# Physical model for the latent heat of fusion

Jozsef Garai

Department of Earth Sciences
Florida International University
University Park, PC-344
Miami, FL 33199
Ph: 305-348-3445
Fax: 305-348-3070
E-mail: jozsef.garai@fiu.edu

**Abstract**

The atomic movement induced on melting has to overcome a viscous drag resistance. It is suggested that the latent heat of fusion supplies the required energy for this physical process. The viscosity model introduced here allows computation of the latent heat from viscosity, molar volume, melting temperature, and atomic mass and diameter. The correlation between these parameters and the latent heat of 14 elements with body and face centered cubic structures was exceptional, with the correlation coefficients of 0.97 and 0.95 respectively.

**1. Introduction**

Despite the enormous importance of the liquid-solid phase transformation there is still no satisfactory theory able to describe this process. One of the key missing elements is an understanding of the latent heat of fusion [$L_f$]. The latent heat of fusion is the energy which has to be supplied to the system in order to complete the phase transformation. This energy is absorbed at constant temperature when the solid transforms to liquid. Since the temperature remains constant, the corresponding entropy change is the difference between the entropy [S] of the two phases.

$$\Delta S = S_{liquid} - S_{solid}, \qquad (1)$$

then the latent heat of fusion is

$$L_f = T_m \Delta S, \qquad (2)$$

where $T_m$ is the melting temperature in degrees of Kelvin. The melting or solidification of a crystalline solid is always a first order transition and heat is usually absorbed or released at the melting temperature (Fig. 1a). There is only one exception to this general rule [1; 2] the transition of solid helium to liquid helium II (Fig. 1b). Not every solid liquid phase transformation requires latent heat for its transition [3; 4]. Non-crystalline solids or glasses do not have a well defined melting temperature and latent heat is not required for the completion of their phase transformation (Fig. 1c). This letter explores the physical process behind the latent heat and presents a new model to explain the different transformations of crystalline, non-crystalline solids, and helium.

## 2. Proposed model

Investigating the fundamental characteristics of the solid and liquid phases, one of the most striking differences between them is the atomic position stability. In solids the atomic positions are well defined while in liquids the position stability is lost. The induced atomic movement at melting most likely will generate viscous resistance. It is suggested that the energy of the latent heat of fusion is utilized to overcome this viscous resistance. This physical explanation of the latent heat is consistent with the zero viscosity of helium II and with the glass transformation. If the latent heat supplies the energy for the viscous drag, then liquid with zero viscosity should not require latent heat for its phase transformation. The transition between solid and liquid helium II requires no energy (Fig. 1b).

When non-crystalline solids solidify, the atomic movements are not restricted to crystal lattice sites as they are in crystalline solids. Instead, there is a continuous reduction in atomic mobility and the energy (latent heat) is released through the whole temperature range of the phase transformation. The lack of definite melting point and latent heat for non-crystalline solids (Fig. 1c) is consistent with the proposed physical explanation.

## 3. Calculations

The thermal velocity of the atoms in liquid is

$$v_{therm} = \sqrt{\frac{kT_m}{m}}, \tag{3}$$

where k is the Boltzmann constant and m is the atomic mass. The maximum velocity difference between two atoms is $2v_{therm}$ while the minimum is zero. It will be assumed that the average velocity difference between the atoms is $v_{therm}$.

For Newtonian liquids the viscosity [η] is defined as

$$\eta = \frac{\tau}{\dot{\varepsilon}}, \tag{4}$$

where $\tau$ is the shear stress, and $\dot{\varepsilon}$ is the strain rate. The strain rate for atoms moving with the velocity difference of $v_{therm}$ can be written as

$$\dot{\varepsilon} = \frac{v_{therm}}{(n+1)d}, \tag{5}$$

where (n+1)d is the distance between the center of the atoms. It was assumed that the atoms are displaced from each other by their atomic diameter [d] and n is the number of the coupled atoms between the two atoms moving with the velocity of $v_{therm}$.

The shear stress is the ratio of the viscous drag force [$F_{vd}$] and the surface of the sheared area [A]

$$\tau = \frac{F_{vd}}{A}. \tag{6}$$

The total sheared area or surface $[A_{mol}]$ in one molar volume $[V_{mol}]$ is estimated as

$$A_{mol} = \frac{V_{mol}}{(n+1)d}. \tag{7}$$

The viscous drag force for one mol of liquid then is

$$F_{vd-mol} = \tau\, A_{mol} = \eta \dot{\varepsilon} A_{mol} = \eta \frac{v_{ther}}{(n+1)d} \frac{V_{mol}}{(n+1)d} = \frac{1}{(n+1)^2} \frac{\eta v_{ther} V_{mol}}{d^2}. \tag{8}$$

The estimated distance between the neighboring potential wells is equivalent to the atomic diameter. The energy $[E_{mol}]$ needed to move all the atoms from one potential well to the next one is

$$E_{mol} = F_{vd-mol} d. \tag{9}$$

If this extra energy is supplied at the melting temperature then the displacement of the atoms becomes possible. The required latent heat for melting is therefore

$$L_f = E_{mol} = \frac{1}{(n+1)^2} \frac{V_{mol}}{d} \eta v_{ther}. \tag{10}$$

The number of coupled atoms should be constant for melts formed from the same crystal structure. Using equation 10 the number of the coupled atoms was calculated

$$n = \sqrt{\frac{\eta v_{ther} V_{mol}}{L_f d}} - 1. \tag{11}$$

The average values and their standard deviations are 1.93 (±0.28), 1.47 (±0.31), and 1.57 (±0.18) for liquids formed from bcc, fcc, and hexagonal close-packed structure respectively. The calculated 1-2 coupled atoms for the different melts seem to be reasonable.

Equation 10 has been tested using experimental data for liquids formed from three different crystal structures. The physical properties of the 17 elements used for the investigation are listed in Table 1. The observed $L_f$ correlates well with the latent heat of fusion as predicted from equation 10 and the physical properties in Table 1 (Fig. 2). The calculated correlation coefficients between all the variables and the latent heat are 0.97 and 0.95 for liquids formed from body centered cubic (bcc) and from face centered cubic (fcc) structures respectively.

## 4. Conclusions

It is proposed that the energy of the latent heat of fusion is required to overcome viscous drag resistance introduced at melting. Assuming that the atoms are moving with their thermal velocity in the liquid, the viscous resistance working against this movement was calculated for melts formed from highly symmetrical packing arrangements. The calculated energies correlate very well with experimentally determined latent heat values.

**Acknowledgement**

I thank to Prof. M. Sukop for reading and commenting the manuscript.

**References**

[1]  C.A. Swenson, Phys. Rev. 79 (1950) 626.
[2]  C.A. Swenson, Phys. Rev. 86 (1952) 870.
[3]  P.G. Debenedetti and F.H. Stillinger, Nature 410 (2001) 259.
[4]  C.A. Angell, K.L. Ngai, G.B. McKenna, P.F. McMillan and S.W. Martin, J. Appl. Phys. 88 (2000) 3113.
[5]  R. F. Brooks, I. Egri, S. Sheetharaman and D. Grant, High Temp.–High Press. 33 (2001) 631.
[6]  Handbook of Chemistry and Physics, CRC Press, Inc., LLC, 1998
[7]  Handbook of Physical Quantities, ed. I. S. Grigoriev and E. Z. Meilikhov, CRC Press, Inc., Boca Raton, 1997

**Table 1**

| Element | Crystal Structure | Atomic Diameter $[10^{-10} m]$ | Liquid Mol. Vol. $[10^{-6} m^3]$ | Viscosity $[10^{-4}$ Pa s$]$ | Latent Heat of Fusion [kJ/mol] | Melting Temp. [K] | Atomic Weight [amu] |
|---|---|---|---|---|---|---|---|
| Li | bcc | 3.04 | 13.47 | 5.66 | 2.33 | 453.69 | 6.941 |
| Na | bcc | 3.72 | 24.77 | 6.87 | 2.60 | 370.96 | 22.990 |
| K | bcc | 4.72 | 47.22 | 4.41 | 2.40 | 336.80 | 39.098 |
| Fe | bcc | 2.52 | 7.938 | 58.0 | 13.81 | 1808.00 | 55.845 |
| Rb | bcc | 4.95 | 57.95 | 5.42 | 2.19 | 312.20 | 85.468 |
| Cs | bcc | 5.31 | 72.11 | 5.98 | 2.09 | 301.55 | 132.905 |
| Ca | bcc | 3.94 | 29.36 | 11.1 | 8.54 | 1112.00 | 40.078 |
| Ar | fcc | 3.84 | 28.17 | 2.80 | 1.21 | 83.78 | 39.948 |
| Al | fcc | 2.86 | 11.29 | 12.9 | 10.71 | 933.52 | 26.982 |
| Ni | fcc | 2.50 | 7.546 | 43.5 | 17.48 | 1726.00 | 58.693 |
| Cu | fcc | 2.56 | 8.003 | 43.2 | 13.26 | 1356.60 | 63.546 |
| Ag | fcc | 2.88 | 11.54 | 38.0 | 11.30 | 1235.08 | 107.868 |
| Au | fcc | 2.88 | 11.40 | 51.3 | 12.55 | 1337.58 | 196.967 |
| Pb | fcc | 3.50 | 19.40 | 26.0 | 4.77 | 600.65 | 207.200 |
| Mg | hcp | 3.20 | 15.34 | 11.0 | 9.04 | 922.00 | 24.305 |
| Co | hcp | 2.50 | 7.684 | 41.5 | 16.2 | 1768.00 | 58.933 |
| Zn | hcp | 2.66 | 8.627 | 33.0 | 7.32 | 692.73 | 65.390 |

Physical parameters of the liquids used for this investigation.
The viscosity of Cu, Ni, and Fe is from Ref. 5
The viscosity of the rest of the metals is from Ref. 6
The rest of the data are from Ref. 7

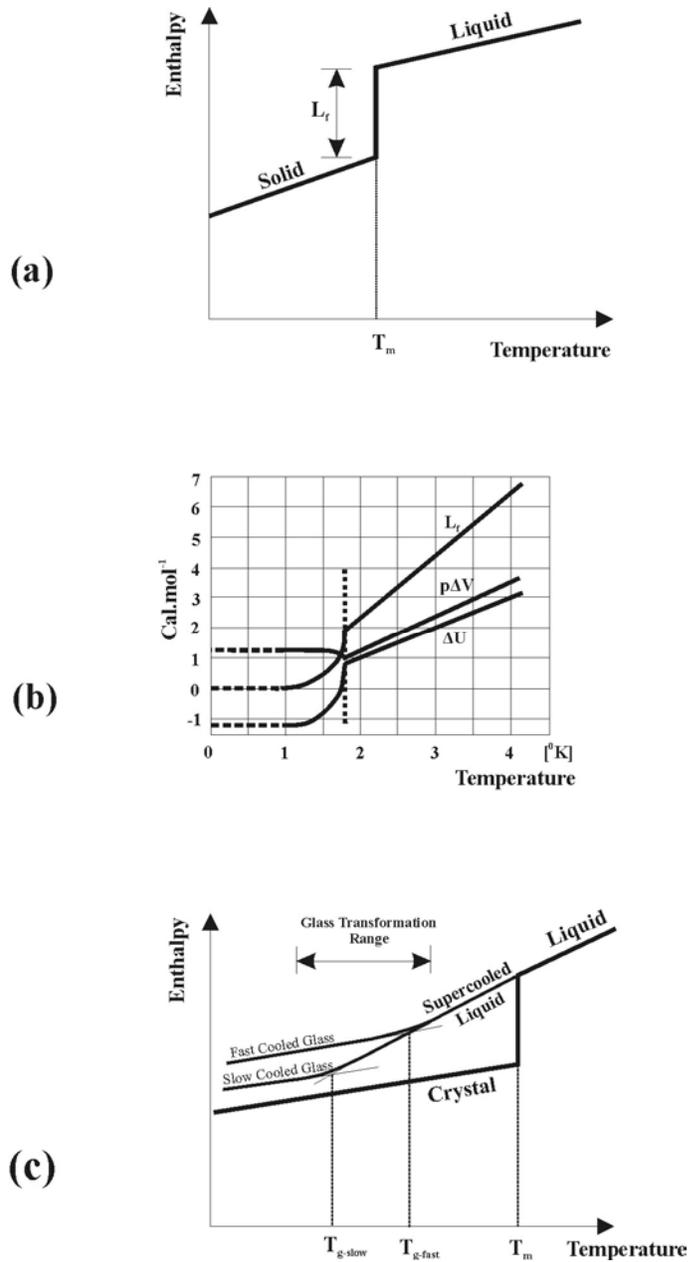

Fig. 1  Melting or solidification.
a.   Crystalline solids
b.   Solid helium-liquid helium II.  The molar latent heat of fusion $L_f$, $p\Delta V$, and the change in internal energy, $\Delta U = L_f - p\Delta V$ for helium from Ref. 2.
c.   Glasses

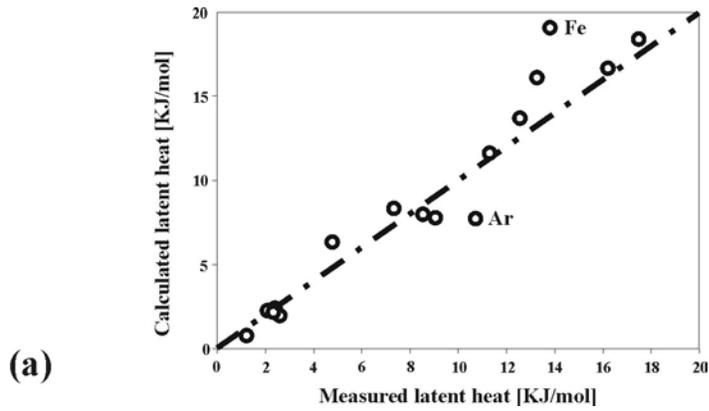

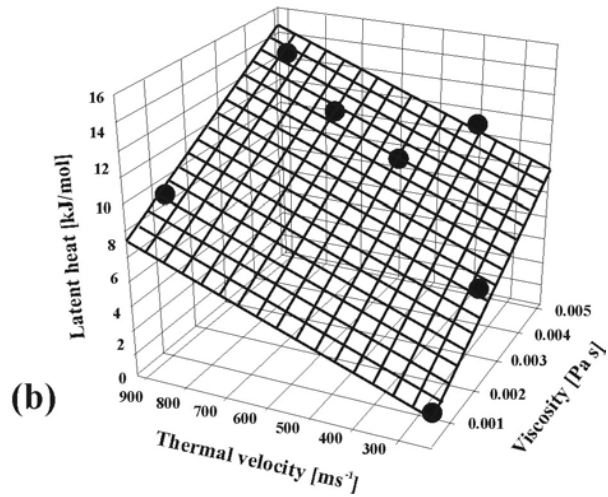

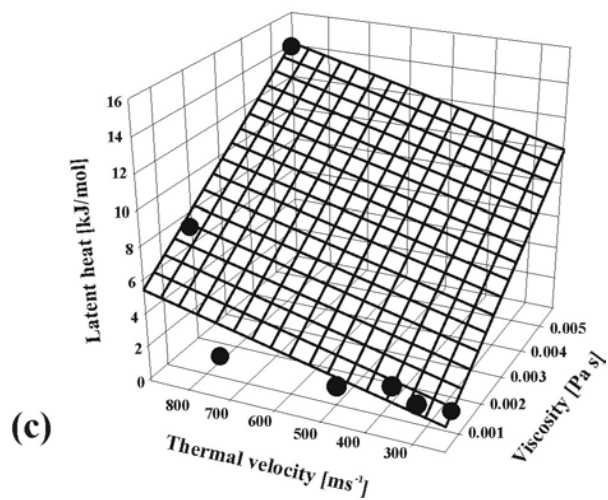

Fig. 2   Correlations between the measured and calculated latent heats and between the measured latent heat and the physical parameters used for the calculation.
a.     Plot of the measured and calculated latent.  The average of the number of the coupled atoms was used for each of the structures.
b.-c.   Correlations between thermal velocity, viscosity and latent heat